\documentclass[twocolumn,superscriptaddress,letter]{revtex4}%
\usepackage{amssymb}
\usepackage{color}
\usepackage{graphicx}
\usepackage{dcolumn}
\usepackage{bm}
\usepackage[header,title,page,titletoc]{appendix}
\usepackage{amsmath}
\usepackage{amsfonts}%
\providecommand{\U}[1]{\protect \rule{.1in}{.1in}}

\begin{document}
\title{Defective Edge States and Anomalous Bulk-Boundary Correspondence in
non-Hermitian Topological Systems}
\author{Xiao-Ran Wang}
\thanks{These authors contributed equally to the work}
\affiliation{Center for Advanced Quantum Studies, Department of Physics, Beijing Normal
University, Beijing 100875, China}
\author{Cui-Xian Guo}
\thanks{These authors contributed equally to the work}
\affiliation{Center for Advanced Quantum Studies, Department of Physics, Beijing Normal
University, Beijing 100875, China}
\author{Su-Peng Kou}
\thanks{Corresponding author}
\email{spkou@bnu.edu.cn}
\affiliation{Center for Advanced Quantum Studies, Department of Physics, Beijing Normal
University, Beijing 100875, China}

\begin{abstract}
Non-Hermitian topological systems show quite different properties as their
Hermitian counterparts. An important, puzzled issue on non-Hermitian
topological systems is the existence of defective edge states beyond usual
bulk-boundary correspondence (BBC) that localize either on the left edge or
the right edge of the one-dimensional system. In this paper, to understand the
existence of the defective edge states, the theory of anomalous bulk-boundary
correspondence (A-BBC) is developed that distinguishes the non-Bloch
bulk-boundary correspondence (NB-BBC) from non-Hermitian skin effect. By using
the one-dimensional non-Hermitian Su-Schrieffer-Heeger model as an example,
the underlying physics of defective edge states is explored. The defective
edge states are physics consequence of boundary exceptional points of
anomalous edge Hamiltonian. In addition, with the help of a\ theorem, the
number anomaly of the edge states in non-Hermitian topological systems become
a mathematic problem under quantitative calculations by identifying the
Abelian/non-Abelian non-Hermitian condition for edge Hamiltonian and verifying
the\emph{\ }deviation of the BBC ratio from $1$. In the future, the theory for
A-BBC can be generalized to higher dimensional non-Hermitian topological
systems (for example, 2D Chern insulator).

\end{abstract}

\pacs{11.30.Er, 75.10.Jm, 64.70.Tg, 03.65.-W}
\maketitle

\textbf{Introduction to non-Hermitian topological systems: }Topological
systems, including topological insulators and topological superconductors are
new types\textbf{ }of\emph{ }exotic quantum phases of matter that have become
the forefront of condensed matter physics for many
years\cite{Kane2010,Qi2011,Ali2012,chi2016,ban2016}. An important topological
property for different types of topological systems is bulk-boundary
correspondence (BBC), i.e., bulk topological invariants that characterize the
topological systems correspond to unique gapless boundary (edge) states.
Recently, non-Hermitian (NH) topological systems have been intensively studied
in both
theory\cite{Rudner2009,Esaki2011,Hu2011,Liang2013,Zhu2014,Lee2016,San2016,Leykam2017,Shen2018,Lieu2018,
Xiong2018,Kawabata2018,Gong2018,Yao2018,YaoWang2018,Yin2018,Kunst2018,KawabataUeda2018,Alvarez2018,
Jiang2018,Ghatak2019,Avila2019,Jin2019,Lee2019,Liu2019,38-1,38,chen-class2019,Edvardsson2019,
Herviou2019,Yokomizo2019,zhouBin2019,Kunst2019,Deng2019,SongWang2019,xi2019,Longhi2019,chen-edge2019}
and
experiments\cite{Zeuner2015,Weimann2017,Xiao2017,Bandres2018,Zhou20182,Cerjan2019,Wang2019,Xiaoxue2019,Helbig2019}%
. The topological properties of NH systems are quite different with their
Hermitian counterparts, including the fractional topological invariant and
defective edge states\cite{Lee2016,Yin2018}, the breakdown of traditional
BBC\cite{Xiong2018,Yao2018,YaoWang2018,Kunst2018,Herviou2019,Yokomizo2019,Kunst2019,Deng2019,Longhi2019}
and NH skin effect\cite{Yao2018,Ghatak2019,Lee2019,SongWang2019,Longhi2019},
and so on. Recently, within the framework of Altland-Zirnbauer (AZ) theory,
the classification of NH systems with topological bands is characterized by
different symmetry-protected topological invariants\cite{Gong2018,38,38-1}.

\textbf{Puzzle about defective edge states}: An important, \emph{puzzled}
issue on NH topological systems is the existence of defective edge states
(DESs) beyond usual BBC. The DES is a zero mode localized either on the left
edge or the right edge of a one-dimensional (1D) finite NH topological system
that is firstly discovered numerically by Lee in 2016. In Ref.\cite{Lee2016},
Lee pointed out that it is the \emph{fractional winding number} that
guarantees the DESs and a new type of BBC exists. However, due to the
existence of non-Hermitian skin effect, the fractional winding number
corresponding to DESs fails. In Ref.\cite{Yao2018}, Yao and Wang pointed out
that it is \emph{non-Bloch topological invariables} rather than fractional
winding number that characterize the topological properties of the NH
topological system. The energy spectra of the NH topological system under open
boundary condition (OPB) may differ from those under periodic boundary
condition (PBC). The new BBC is called non-Bloch BBC or \emph{NB-BBC} for
short. With the help of NB-BBC, people can obtain the correct topological
phase diagram and know the condition whether the edge states exist or
not\cite{Yao2018}. Instead of single DES, according to NB-BBC, two end states
are predicted to separately locate at the two opposite sides, or both locate
at one (left or right) side.

So, the situation becomes confusing: according to numerical calculations,
there exist DESs in NH topological systems. To accurately predict the
existence of single DES, one should obtain the topological phase diagram to
know whether the edge states exist and check the number of the edge states to
know whether the edge states become defective. However, neither the fractional
winding number nor the non-Bloch topological invariables accurately predict
the existence of single DES. There is a puzzle: \emph{how to understand the
existence of the DESs}?

In this paper, we try to resolve this puzzle after answering the following questions:

\begin{enumerate}
\item What's the \emph{underlying }physics of the DESs?

\item How to \emph{accurately }characterize the DESs?

\item Are the DESs \emph{stable} in thermodynamic limit?

\item Does there exist \emph{universal} features of DESs?
\end{enumerate}

The key point of the answers is which the number of the edge states becomes
anomalous due to the existence of DESs that distinguishes NB-BBC with
phase-boundary anomaly. We call the anomalous BBC with the number anomaly of
the edge states \emph{A-BBC} for short.

\textbf{Quantitative description for anomalous bulk-boundary correspondence}:
Firstly, we consider the boundary physics for an arbitrary 1D finite NH
topological system, of which the Hamiltonian is $\hat{H}_{\mathrm{NH}}$
($\hat{H}_{\mathrm{NH}}\neq \hat{H}_{\mathrm{NH}}^{\dagger}$). From the
classification of NH topological systems\cite{38,38-1}, in topological phase,
the topological invariant $\mathcal{Z}$ may guarantee $\mathcal{C}%
_{\mathrm{finite}}$ edge states in 1D. The \emph{biorthogonal set} for the
quantum states of the boundary/edge modes is defined by $|\mathrm{b}%
_{k}\rangle$ and $|\mathrm{B}_{k}\rangle,$ i.e., $\hat{H}_{\mathrm{NH}%
}|\mathrm{b}_{k}\rangle=E_{k}|\mathrm{b}_{k}\rangle$, $\hat{H}_{\mathrm{NH}%
}^{\dagger}|\mathrm{B}_{k}\rangle=(E_{k})^{\ast}|\mathrm{B}_{k}\rangle,$ and
$\langle \mathrm{B}_{k}|\mathrm{b}_{k}\rangle=1$ ($k=1,...,\mathcal{C}%
_{\mathrm{finite}}$ is state index)\cite{mi}. The BBC for finite NH
topological systems is denoted by $\mathcal{C}_{\mathrm{finite}}=2\mathcal{Z}%
$. A-BBC is \emph{the number anomaly of the edge states}, i.e., the number of
the edge states is not to be $2\mathcal{Z}$ even in thermodynamic limit,
\begin{equation}
\mathcal{C}_{\mathrm{finite}}\neq2\mathcal{Z}.
\end{equation}
A special case of A-BBC is $\mathcal{C}_{\mathrm{finite}}=\mathcal{Z}$ that
indicates the existence of a DES. We emphasize that the DES may have a
distribution on both edges.

In this paper, we focus on 1D finite NH topological system with $\mathcal{Z}%
=1$. To quantitatively characterize the edge physics for a 1D finite NH
topological system, we introduce an \emph{effective} \emph{edge Hamiltonian},
\begin{equation}
\mathcal{\hat{H}}_{\mathrm{edge}}=\left(
\begin{array}
[c]{cc}%
h_{11} & h_{12}\\
h_{21} & h_{22}%
\end{array}
\right)
\end{equation}
where $h_{IJ}=\left \langle \mathrm{b}^{I}\right \vert \hat{H}_{\mathrm{NH}%
}\left \vert \mathrm{b}^{J}\right \rangle ,$ $I,J=1,2$. $(%
\begin{array}
[c]{c}%
\left \vert \mathrm{b}^{1}\right \rangle \\
\left \vert \mathrm{b}^{2}\right \rangle
\end{array}
)$ are the basis of the edge states. With the help of $\mathcal{\hat{H}%
}_{\mathrm{edge}}$, the boundary phase diagram of the NH topological systems
can be obtained straightforwardly.

To develop a clear quantitative description for the DESs and the corresponding
A-BBC, we introduce three new concepts:

\emph{Definition 1 -- Abelian/non-Abelian NH condition}: \emph{A NH
Hamiltonian }$\hat{H}$\emph{ (}$\hat{H}\neq(\hat{H})^{\dagger}$\emph{) can be
written into }$\hat{H}=\hat{H}_{h}+\hat{H}_{a}$\emph{ where }$\hat{H}%
_{h}=\frac{1}{2}(\hat{H}+\hat{H}^{\dagger})$ and $\hat{H}_{a}=\frac{1}{2}%
(\hat{H}-\hat{H}^{\dagger})$\emph{ are the Hermitian part and anti-Hermitian
part, respectively. A Hamiltonian }$\hat{H}$\emph{ is Abelian NH if }$[\hat
{H}_{h},\hat{H}_{a}]=0$\emph{;} \emph{A Hamiltonian }$\hat{H}$\emph{ is
non-Abelian NH if }$[\hat{H}_{h},\hat{H}_{a}]\neq0$\emph{. }When
$\mathcal{\hat{H}}_{\mathrm{edge}}$ obeys non-Abelian NH, we call it
anomalous, of which the eigenstates ($|\psi_{+}\rangle$ and $|\psi_{-}\rangle
$) coalesce. Consequently, according to $\mathcal{C}_{\mathrm{finite}}%
\neq2\mathcal{Z},$ the BBC becomes anomalous.

\emph{Definition 2 -- State similarity of edge states}: \emph{The state
similarity for two edge states }$|\psi_{+}\rangle$ and $|\psi_{-}\rangle
$\emph{ is }$|\langle \psi_{+}|\psi_{-}\rangle|$\emph{. Here, }$|\psi_{\pm
}\rangle$\emph{ is satisfied self-normalization condition, i.e., }%
$|\langle \psi_{\pm}|\psi_{\pm}\rangle|\equiv1$\emph{.}

\emph{Definition 3 -- BBC ratio}: \emph{The BBC ratio is defined by }%
$\Upsilon_{\mathrm{BBC}}=1-|\langle \psi_{+}|\psi_{-}\rangle|/2=$\emph{ where
}$\mathcal{C}_{\mathrm{finite}}=2\mathcal{Z}\cdot \Upsilon_{\mathrm{BBC}}%
$\emph{ is the total number of edge states for a finite NH system. }%
$|\psi_{\pm}\rangle$\emph{ are the two edge states. }$\Upsilon_{\mathrm{BBC}}%
$\ is a quantity that characterizes number anomaly of the edge states.

If $\Upsilon_{\mathrm{BBC}}$ is $1$, there exists the usual bulk-boundary
correspondence; if $\Upsilon_{\mathrm{BBC}}$ is smaller than $1$, there exists
A-BBC. And if $\Upsilon_{\mathrm{BBC}}$ is equal to $1/2$, there exists DES.
The existence of DESs and A-BBC in NH topological systems becomes a well
defined mathematic problem under quantitative calculations. The detailed proof
of this statement is given in supplementary materials.

\textbf{The defective edge states as boundary exceptional points in
nonreciprocal Su-Schrieffer-Heeger model:} We take 1D nonreciprocal
Su-Schrieffer-Heeger (SSH) model as an example to explore the underlying
physics of A-BBC and DESs. The Bloch Hamiltonian for a nonreciprocal SSH model
with $N$ pairs of lattice sites under PBC is given by
\begin{equation}
\hat{H}_{\mathrm{NH-SSH}}=\left(  t_{1}+t_{2}\cos k\right)  \sigma_{x}+\left(
t_{2}\sin k+i\gamma \right)  \sigma_{y}.
\end{equation}
$t_{1}$ and $t_{2}$ describe the intra-cell and inter-cell hopping strengths,
respectively. $\gamma$ describes the unequal intra-cell hoppings. $k$ is
wave-vector. $k$, $t_{1},$ $t_{2}$, $\gamma$ are all real. $\sigma_{i}$'s are
the Pauli matrices acting on the (\textrm{A} or \textrm{B}) sublattice
subspace. In this paper, we set $t_{2}=1$. For this case, there exists chiral
symmetry, $\sigma_{z}\hat{H}_{\mathrm{NH-SSH}}\left(  k\right)  \sigma
_{z}=-\hat{H}_{\mathrm{NH-SSH}}\left(  k\right)  $.

Firstly, we consider the Hermitian SSH model with $\gamma=0$. Now, the
topological invariant is a winding number $w=\frac{1}{2\pi}\int_{-\pi}^{\pi
}\partial_{k}\phi(k)\cdot dk$ where $\phi(k)=\tan^{-1}(d_{y}/d_{x})$ with
$d_{y}=t_{2}\sin k$ and $d_{x}=t_{1}+t_{2}\cos k$. There are two phases for
the global bulk phase diagram, topological phase with $w=1$ in the region of
$\left \vert t_{1}\right \vert <\left \vert t_{2}\right \vert $ and trivial phase
with $w=0$ in the region of $\left \vert t_{1}\right \vert >\left \vert
t_{2}\right \vert $. In topological phase, the basis of the edge states is
given by\cite{si}, $(%
\begin{array}
[c]{c}%
\left \vert \mathrm{b}^{1}\right \rangle \\
\left \vert \mathrm{b}^{2}\right \rangle
\end{array}
),$ of which the wave-function is $\left \vert \mathrm{b}^{1}\right \rangle
=\frac{1}{\mathcal{N}}\sum_{n=1}^{N}(-\frac{t_{1}}{t_{2}})^{n-1}%
|n\rangle \otimes(%
\begin{array}
[c]{c}%
1\\
0
\end{array}
)$ or $\left \vert \mathrm{b}^{2}\right \rangle =\frac{1}{\mathcal{N}}\sum
_{n=0}^{N-1}(-\frac{t_{1}}{t_{2}})^{n}|N-n\rangle \otimes(%
\begin{array}
[c]{c}%
0\\
1
\end{array}
)$ where $\mathcal{N}=\sqrt{(1-(\frac{t_{1}}{t_{2}})^{2N})/(1-(\frac{t_{1}%
}{t_{2}})^{2})}$ is normalization factor, and $(%
\begin{array}
[c]{c}%
1\\
0
\end{array}
)$ and $(%
\begin{array}
[c]{c}%
0\\
1
\end{array}
)$ denote the state vectors of two-sublattices. The orthogonal condition for
the two edge states is $\langle \mathrm{b}^{1}|\mathrm{b}^{2}\rangle=0$. To
characterize the dynamics of the two edge states of finite NH SSH model, we
derive the edge Hamiltonian, $\mathcal{\hat{H}}_{\mathrm{edge}}=\left(
\begin{array}
[c]{cc}%
h_{11} & h_{12}\\
h_{21} & h_{22}%
\end{array}
\right)  $ where $h_{IJ}=\left \langle \mathrm{b}^{I}\right \vert \hat
{H}_{\mathrm{NH-SSH}}\left \vert \mathrm{b}^{J}\right \rangle ,$ $I,J=1,2$. For
this case, we have $h_{11}=-h_{22}=0,$ $h_{12}=h_{21}=\Delta_{0}=\frac
{(t_{1}^{2}-t_{2}^{2})}{t_{2}}(-\frac{t_{1}}{t_{2}})^{N}$ and obtain an
effective edge Hamiltonian for Hermitian SSH model
\begin{equation}
\mathcal{\hat{H}}_{\mathrm{edge}}=\Delta_{0}\cdot \tau^{x}%
\end{equation}
where $\tau^{i}$'s are the Pauli matrices acting on the subspace of two edge states.

When consider the unequal intra-cell hoppings ($\gamma \neq0$), the DESs
appear. Correspondingly, the NB-BBC becomes anomalous.

Under OBC, to deal with NH skin effect\cite{Yao2018,YaoWang2018}, we do a
similar-transformation, i.e., $\bar{H}_{\mathrm{NH-SSH}}=(\mathcal{\hat{S}%
}_{\mathrm{NHP}})^{-1}\hat{H}_{\mathrm{NH-SSH}}\mathcal{\hat{S}}%
_{\mathrm{NHP}}=(\bar{t}_{1}+\bar{t}_{2}\cos k)\sigma_{x}+\bar{t}_{2}\sin
k\sigma_{y}$ where $\mathcal{\hat{S}}_{\mathrm{NHP}}$ is the operation for
similar transformation. Under $\mathcal{\hat{S}}_{\mathrm{NHP}},$ we remove
the imaginary wave-vector, i.e., $\left \vert \mathrm{\psi}(k)\right \rangle
\rightarrow \left \vert \mathrm{\psi}(k)\right \rangle ^{\prime}=(\mathcal{\hat
{S}}_{\mathrm{NHP}})^{-1}\left \vert \mathrm{\psi}(k-iq_{0})\right \rangle $ by
doing NH transformation $U_{\mathrm{NHP}}=(%
\begin{array}
[c]{cc}%
1 & 0\\
0 & e^{-q_{0}}%
\end{array}
)$ and site-dependent scaling transformation $|n\rangle \rightarrow
|n\rangle^{\prime}=e^{-q_{0}(n-1)}|n\rangle$ ($n$ denotes the cell number).
$q_{0}$ is the real value of imaginary wave-vector, $e^{q_{0}}=\sqrt
{\frac{t_{1}-\gamma}{t_{1}+\gamma}}$ Consequently, the effective hopping
parameters become $\bar{t}_{1}=\sqrt{(t_{1}-\gamma)(t_{1}+\gamma)},\quad$and
$\bar{t}_{2}=t_{2}.$ Now, the topological transition occurs at $\left \vert
\bar{t}_{1}\right \vert =\left \vert \bar{t}_{2}\right \vert .$

In the topological phase $\left \vert \bar{t}_{1}\right \vert <\left \vert
\bar{t}_{2}\right \vert $, the edge states are protected by the non-Bloch
topological invariants determined by $\bar{H}_{\mathrm{NH-SSH}}$%
\cite{Yao2018,YaoWang2018}, $w=\frac{1}{2\pi}\int_{-\pi}^{\pi}\partial
\bar{\phi}(k)dk$, where $\bar{\phi}(k)=\tan^{-1}(\bar{d}_{y}/\bar{d}_{x})$ and
$\bar{d}_{x}=\bar{t}_{1}+\bar{t}_{2}\cos k$, $\bar{d}_{y}=\bar{t}_{2}\sin k$.
Due to $w=1$, based on biorthogonal set, the basis of the edge states is given
by\cite{Lee2016,zhouBin2019,chen-edge2019}, $(%
\begin{array}
[c]{c}%
\left \vert \mathrm{b}^{1}\right \rangle \\
\left \vert \mathrm{b}^{2}\right \rangle
\end{array}
)$, of which the wave-function is $\left \vert \mathrm{b}^{1}\right \rangle
=\frac{1}{\mathcal{\bar{N}}}\sum_{n=1}^{N}(-\frac{\bar{t}_{1}}{\bar{t}_{2}%
})^{n-1}e^{q_{0}(n-1)}|n\rangle \otimes(%
\begin{array}
[c]{c}%
1\\
0
\end{array}
),$ or $\left \vert \mathrm{b}^{2}\right \rangle =\frac{1}{\mathcal{\bar{N}}%
}\sum_{n=0}^{N-1}(-\frac{\bar{t}_{1}}{\bar{t}_{2}})^{n}e^{-q_{0}n}%
|N-n\rangle \otimes(%
\begin{array}
[c]{c}%
0\\
1
\end{array}
)$ where $\mathcal{\bar{N}}=\mathcal{N}(t_{1}\rightarrow \bar{t}_{1}%
,t_{2}\rightarrow \bar{t}_{2})=\sqrt{(1-(\frac{\bar{t}_{1}}{\bar{t}_{2}}%
)^{2N})/(1-(\frac{\bar{t}_{1}}{\bar{t}_{2}})^{2})}$ is normalization factor.
Because the system with $\bar{t}_{1}=\sqrt{(t_{1}-\gamma)(t_{1}+\gamma)}=0$ is
singularity that corresponds to EPs of all bulk states, in this paper we focus
on the case of $\bar{t}_{1}\neq0$.

\begin{figure}[ptb]
\includegraphics[clip,width=0.5\textwidth]{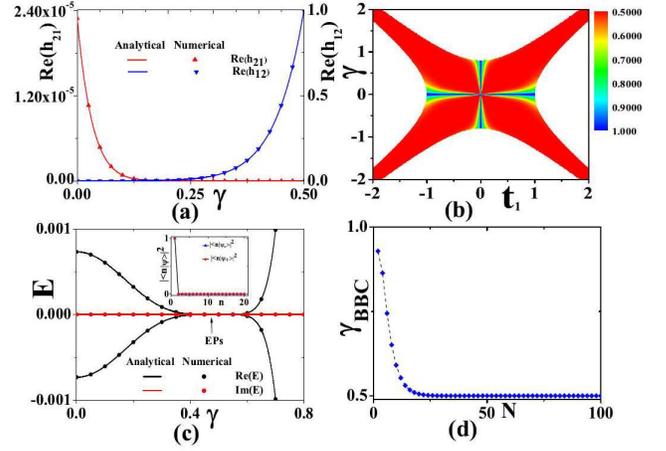}\caption{(Color online)
(a) The numerical results and the analytic results for two matrix elements
$h_{12},$ $h_{21}$ of the effective edge Hamiltonian $\mathcal{\hat{H}%
}_{\mathrm{edge}}$ for the case of $t_{1}=0.5$ and $N=15;$ (b) The BBC ratio
$\Upsilon_{\mathrm{BBC}}$. The red region corresponds to boundary EPs. In red
region, the bulk-boundary correspondence ratio $\Upsilon_{\mathrm{BBC}}$ is
about $0.5$. The number of unit cells is $N=10$; (c) The numerical results and
the analytic results for energy level of the edge states for the case of
$t_{1}=0.5$ and $N=10$. The inset is the wave-function of the edge states near
exceptional point with $\gamma=0.49$. Here, $n$ denotes lattice site; (d) The
BBC ratio via lattice number $N$ for the case of $\gamma=0.7,$ $t_{1}=0.2$.
The result indicates that in thermodynamic limit, the BBC ratio is $1/2$. The
BBC becomes anomalous.}%
\end{figure}

Under similar transformation $t_{1}\rightarrow \bar{t}_{1},$ $t_{2}%
\rightarrow \bar{t}_{2}$, the effective edge Hamiltonian becomes
\begin{align}
\mathcal{\hat{H}}_{\mathrm{edge}}(t_{1},t_{2})  &  =\Delta_{0}\cdot \tau^{x}\\
&  \rightarrow \mathcal{\bar{H}}_{\mathrm{edge}}(t_{1}\rightarrow \bar{t}%
_{1},t_{2}\rightarrow \bar{t}_{2})=\bar{\Delta}\cdot \tau^{x}\nonumber
\end{align}
where $\bar{\Delta}=\frac{(\bar{t}_{1}^{2}-\bar{t}_{2}^{2})}{\bar{t}_{2}%
}(-\frac{\bar{t}_{1}}{\bar{t}_{2}})^{N}.$ The energy levels are correct that
are $E_{\pm}=\pm \bar{\Delta}.$ However, although energy levels are correct,
the detailed numerically calculations \emph{cannot} support this effective
edge Hamiltonian $\mathcal{\bar{H}}_{\mathrm{edge}}$, for example, in
numerical results $h_{12}=\left \langle \mathrm{b}^{1}\right \vert \hat
{H}_{\mathrm{NH-SSH}}\left \vert \mathrm{b}^{2}\right \rangle \neq
h_{21}=\left \langle \mathrm{b}^{2}\right \vert \hat{H}_{\mathrm{NH-SSH}%
}\left \vert \mathrm{b}^{1}\right \rangle $. What's wrong with $\mathcal{\bar
{H}}_{\mathrm{edge}}$? The mismatch comes from overlooking NH polarization
effect on the two edge states.

To derive the correct result, we must consider an \emph{additional} NH
boundary polarization operation on edge modes, $\tau^{x}\rightarrow
U_{\mathrm{edge}}^{-1}\tau^{x}U_{\mathrm{edge}}=\tau^{x}\cosh(q_{0}%
N)-i\tau^{y}\sinh(q_{0}N)$ where
\begin{equation}
U_{\mathrm{edge}}=(%
\begin{array}
[c]{cc}%
1 & 0\\
0 & e^{-q_{0}N}%
\end{array}
).
\end{equation}
After considering $U_{\mathrm{edge}}$, the effective spin model $\mathcal{\bar
{H}}_{\mathrm{edge}}$ becomes \emph{anomalous}, i.e.,
\begin{align}
\mathcal{\breve{H}}_{\mathrm{eff}}  &  =(U_{\mathrm{edge}})^{-1}%
\mathcal{\bar{H}}_{\mathrm{edge}}(U_{\mathrm{edge}})\nonumber \\
&  =(%
\begin{array}
[c]{cc}%
0 & \bar{\Delta}e^{-q_{0}N}\\
\bar{\Delta}e^{q_{0}N} & 0
\end{array}
)=\bar{\Delta}^{x}\tau^{x}-i\bar{\Delta}^{y}\tau^{y}%
\end{align}
where $\bar{\Delta}^{x}=\bar{\Delta}\cosh(q_{0}N),$ $\bar{\Delta}^{y}%
=\bar{\Delta}\sinh(q_{0}N).$ Under $U_{\mathrm{edge}}$, the energy levels
doesn't change, $E_{\pm}=\pm \bar{\Delta}.$ In Fig.1(a), we also numerically
calculate the two matrix elements for the case of $N=15$ and eventually find
the consistence between the numerical results and analytical results.

There exists emergent chiral symmetry and effective $\mathcal{PT}$-symmetry
for the edge Hamiltonian $\mathcal{\breve{H}}_{\mathrm{eff}}$%
\cite{Bender98,Bender02,Bender07}, i.e.,
\begin{equation}
\tau_{z}\mathcal{\breve{H}}_{\mathrm{eff}}\left(  k\right)  \tau
_{z}=-\mathcal{\breve{H}}_{\mathrm{eff}},
\end{equation}
and
\begin{equation}
\left[  \mathcal{P}_{\mathrm{eff}}\mathcal{T},\text{ }\mathcal{\breve{H}%
}_{\mathrm{eff}}\right]  =0,\text{ }\mathcal{P}_{\mathrm{eff}}=\tau^{x}.
\end{equation}
This anomalous edge Hamiltonian $\mathcal{\breve{H}}_{\mathrm{eff}}$ obeys
non-Abelian NH condition. The condition for boundary EPs is obtained as
$\left \vert \bar{\Delta}\right \vert =0.$ The solution for this equation is
$N\rightarrow \infty$ or $\left \vert t_{1}\right \vert =\left \vert
\gamma \right \vert .$ As shown in Fig.1(b). the red region corresponds to
boundary EPs. Thus, the two edge states merge into one at boundary EPs
$\left \vert t_{1}\right \vert =\left \vert \gamma \right \vert $ or $N\rightarrow
\infty$, of which the wave-function is $\left \vert \psi_{+}\right \rangle
=\left \vert \psi_{-}\right \rangle =\left \vert \mathrm{b}^{1}\right \rangle $ or
$\left \vert \mathrm{b}^{2}\right \rangle $. In thermodynamic limit, the edge
state for $E=0$ becomes defective and is localized either on the left edge or
the right edge.

Let us discuss the A-BBC in this condition. The figure of $\Upsilon
_{\mathrm{BBC}}$ in Fig.1(b) is plotted for NH topological system with finite
size $N=10$. In thermodynamic limit $N\rightarrow \infty$, for the case of
$t_{1}\neq0$ and $\gamma \neq0$, the anomalous edge Hamiltonian becomes
\begin{equation}
\mathcal{\breve{H}}_{\mathrm{eff}}\rightarrow \bar{\Delta}\tau^{\pm},\text{
}(\tau^{+}=\tau^{x}\pm i\tau^{y})
\end{equation}
that also obeys non-Abelian NH condition. Now, we have the state similarity
$|\langle \psi_{+}|\psi_{-}\rangle|=\tanh(2q_{0}N)$ to be $1$ and the BBC ratio
$\Upsilon_{\mathrm{BBC}}$ turns to $\frac{1}{2}.$ In particular, as shown in
Fig.1(d), the BBC for the NH topological system in thermodynamic limit
($t_{1}\neq0$ and $\gamma \neq0$) is anomalous! For the case of $t_{1}=0$ and
$\gamma=0$, the edge Hamiltonian $\mathcal{\breve{H}}_{\mathrm{eff}}$ becomes
usual and the corresponding $\Upsilon_{\mathrm{BBC}}$ turns to $1$.

\textbf{Stability of defective edge states in thermodynamic limit:} To examine
the stability of the DESs, we add an additional imaginary staggered potential
$i\varepsilon \sigma_{z}$ on $\hat{H}_{\mathrm{NH-SSH}}$. Using a\ similar
approach, the anomalous edge Hamiltonian is obtained as
\begin{equation}
\mathcal{\breve{H}}_{\mathrm{eff}}=\bar{\Delta}^{x}\tau^{x}-i\bar{\Delta}%
^{y}\tau^{y}+i\varepsilon \tau^{z}%
\end{equation}
where $\bar{\Delta}^{x}=\bar{\Delta}\cosh(q_{0}N),$ $\bar{\Delta}^{y}%
=\bar{\Delta}\sinh(q_{0}N),$ $e^{q_{0}}=\sqrt{\frac{t_{1}-\gamma}{t_{1}%
+\gamma}},$ $\bar{\Delta}=\frac{(t_{1}^{2}-t_{2}^{2}-\gamma^{2})}{t_{2}%
}(-\frac{t_{1}^{2}-\gamma^{2}}{t_{2}^{2}})^{N/2}$. A spontaneous (effective)
$\mathcal{PT}$-symmetry-breaking transition occurs at the boundary EPs
$\left \vert \bar{\Delta}\right \vert =\left \vert \varepsilon \right \vert $. For
a finite system, the DES for boundary EPs described by $\left \vert \psi
_{+}\right \rangle =\left \vert \psi_{-}\right \rangle =\frac{1}{\sqrt
{1+e^{2q_{0}N}}}(\left \vert \mathrm{b}^{1}\right \rangle -ie^{q_{0}N}\left \vert
\mathrm{b}^{2}\right \rangle ).$ This is DES without chiral symmetry that is no
more localized either on the left edge or on the right edge.

In thermodynamic limit, according to $\bar{\Delta}\rightarrow0,$ for the case
of $\varepsilon \neq0,$ the edge Hamiltonian obeys Abelian NH condition
$\mathcal{\hat{H}}_{\mathrm{edge}}\rightarrow i\varepsilon \tau^{z}.$ The BBC
ratio $\Upsilon_{\mathrm{BBC}}$ turns into $1$ and the BBC becomes usual. That
means, in thermodynamic limit, arbitrary imaginary staggered potential
breaking chiral symmetry overcomes the effect from NH boundary polarization
($\left \vert \varepsilon \right \vert \gg \left \vert \bar{\Delta}\right \vert
\sim \left \vert \frac{t_{1}^{2}-\gamma^{2}}{t_{2}^{2}}\right \vert
^{N/2}\rightarrow0$) and moves the edge Hamiltonian away from boundary EPs. As
a result, in thermodynamic limit, for a general NH SSH model without the
protection of chiral symmetry, the DES is \emph{unstable}.

\textbf{A-BBC theorem for number anomaly of the edge states in a general
non-Hermitian topological system}: To develop general formula, we consider an
arbitrary 1D NH topological system with topological invariant $\mathcal{Z},$
of which the Hamiltonian is $\hat{H}_{\mathrm{NH}}$. To explore the universal
features for A-BBC from DESs, we prove the following A-BBC theorem.

\emph{A-BBC theorem}: \emph{The usual BBC is satisfied if the boundary/edge
Hamiltonian }$\mathcal{\hat{H}}_{\mathrm{edge}}\mathcal{\ }$\emph{(}%
$\mathcal{\hat{H}}_{\mathrm{edge}}\neq(\mathcal{\hat{H}}_{\mathrm{edge}%
})^{\dagger}$\emph{) for a NH topological system obeys Abelian NH condition,
}$[\mathcal{\hat{H}}_{\mathrm{edge},h},\mathcal{\hat{H}}_{\mathrm{edge},a}%
]=0$\emph{. Now, }$\Upsilon_{\mathrm{BBC}}$\emph{ is }$1$\emph{; On the
contrary, the BBC becomes anomalous if the }$\mathcal{\hat{H}}_{\mathrm{edge}%
}$\emph{ obeys the non-Abelian NH condition, }$[\mathcal{\hat{H}%
}_{\mathrm{edge},h},\mathcal{\hat{H}}_{\mathrm{edge},a}]\neq0$\emph{. Now,
}$\Upsilon_{\mathrm{BBC}}$\emph{ is smaller than }$1$\emph{. Here,
}$\mathcal{\hat{H}}_{\mathrm{edge},h}=\frac{1}{2}(\mathcal{\hat{H}%
}_{\mathrm{edge}}+(\mathcal{\hat{H}}_{\mathrm{edge}})^{\dagger})$ and
$\mathcal{\hat{H}}_{\mathrm{edge},a}=\frac{1}{2}(\mathcal{\hat{H}%
}_{\mathrm{edge}}-(\mathcal{\hat{H}}_{\mathrm{edge}})^{\dagger}).$\emph{\ }In
supplementary materials, we show the detailed proof of BBC theorem.\emph{ }

A special case of A-BBC is $\Upsilon_{\mathrm{BBC}}=1/2$ that corresponds to
boundary EPs with the DES. Now, the anomalous edge Hamiltonian $\hat
{H}_{\mathrm{edge}}$ ($\hat{H}_{\mathrm{edge}}\neq(\hat{H}_{\mathrm{edge}%
})^{\dagger}$) obeys the following conditions, $\{ \hat{H}_{\mathrm{edge}%
,h},\hat{H}_{\mathrm{edge},a}\}=0$ and $\hat{H}_{\mathrm{edge},h}=\hat
{T}(i\cdot \hat{H}_{\mathrm{edge},a})\hat{T}^{-1}$ (here, $\hat{T}$ is an
unitary Hermitian matrix).

\textbf{Conclusion and discussion}: In the end, we draw a brief conclusion. In
this paper, the puzzle "\emph{how to understand the existence of the DESs}" is
resolved. The theory of A-BBC for DESs is developed that distinguishes the
NB-BBC from NH skin effect. A rigorous theorem -- A-BBC theorem is proved.
With the help of this theorem, the A-BBC of number anomaly of the edge states
is identified by verifying the NH condition of effective edge Hamiltonian
$\mathcal{\hat{H}}_{\mathrm{edge}}.$ A special case of $\Upsilon
_{\mathrm{BBC}}=1/2$ corresponds to boundary EPs and there exists a DES with
$E=0$. For the NH SSH model with chiral symmetry, unequal\emph{ }intra-cell
hoppings cause NH boundary polarization that drives the edge states to\emph{
}boundary EPs. That means that there is \emph{no} intrinsic relationship
between the DES with zero energy and the fractional winding number for the
bulk state under PBC\cite{Lee2016,Yin2018}. The NB-BBC from NH skin effect for
bulk states is relevant to the correct topological phase diagram but
\emph{not} the defective zero edge modes (nor the number anomaly of the edge
states)\cite{Yao2018,YaoWang2018}. In addition, we emphasize that the
quantitative theory for A-BBC of number anomaly of the edge states can be
generalized to various types of NH topological systems, for example, 1D NH SSH
model with next-next nearest neighbour hoppings, 1D NH topological
superconductors and two dimensional NH topological insulators. These issues
will be presented in future.

\acknowledgments This work is supported by NSFC Grant No. 11674026, 11974053.
We thank Z. Wang, S. Chen, W. Yi\ for helpful discussion.

\end{document}